\documentclass[aip, jcp, preprint]{revtex4-1}

\usepackage{amsfonts, amsmath, amssymb}
\usepackage{amscd, latexsym}
\usepackage{mathrsfs}
\usepackage{epsfig} 
\usepackage[]{graphicx} 
\usepackage{subfig}  
\usepackage{epstopdf} 
\usepackage{bm}
\usepackage{color}
\usepackage{natbib}

\begin{document}

\title{Resonant Energy Transfer Under the Influence of the Evanescent Field from the Metal}

\author{Amrit Poudel}
\thanks{Contributed equally to this work}
\affiliation{\mbox{Department of Chemistry, Northwestern University,}\\
\mbox{2145 Sheridan Road, Evanston, IL 60208, USA}}

\author{Xin Chen}
\thanks{Contributed equally to this work}
\email[E-mail:]{xin.chen.nj@xjtu.edu.cn}
\affiliation{\mbox{Center of Nanomaterials for Renewable Energy, }\\	
\mbox{State Key Laboratory of Electrical Insulation and Power Equipment,}\\
\mbox{School of Electrical Engineering, Xi'an Jiaotong University, Xi'an, China}}

\author{Mark A. Ratner}
\email[E-mail:]{ratner@northwestern.edu}
\affiliation{\mbox{Department of Chemistry, Northwestern University,}\\
\mbox{2145 Sheridan Road, Evanston, IL 60208, USA}}

\begin{abstract}
We present a quantum framework based on a density matrix of a dimer system to investigate the quantum dynamics of excitation energy transfer (EET) in the presence of  the evanescent field from the metal and the phonon bath. Due to the spatial correlation of the electric field in the vicinity of the metal, the spectral density of the evanescent field is similar to that of a shared phonon bath. However, the EET dynamics  under the influence of the evanescent field is an open and a new problem. Here we use a thin metallic film to investigate the effect of the evanescent field on the excitation energy transfer in a dimer system based on a density matrix approach. Our results indicate that a thin metallic film enhances the energy transfer rate at the expense of absorbing energy during the process. Since the spectral density of the evanescent field is affected by the geometry of the medium and the distance of a dimer system from the medium, our results demonstrate the possibility to tune EET based on material geometry and distances.  Our model also serves as an expansion to quantum heat engine model and provides a framework to investigate the EET  in light harvesting molecular networks under the influence of the evanescent field.
\end{abstract}

\maketitle
\section{Introduction}
There is a growing research effort to utilize the electromagnetic environment like photonics crystals, plasmonic nanostructures or dielectric cavities to control and modify the electronic excitation, relaxation and emission processes~\cite{joan,roy,inada}. Recently, a number of studies have been reported that couple exciton to surface polariton leading to optical nonlinearity~\cite{fofang}, induced transparency~\cite{delacy} and topologically protected edge states~\cite{joel}. Indeed, a recent experiment~\cite{orgiu} has taken this a step further by demonstrating that the photonics crystal can be used to enhance the conductivity of the organic semiconductor~\cite{orgiu}. In addition to these studies, the incoherent light source as a photo-thermal reservoir has been used to study the light-matter interaction in molecular systems.  The quantum heat engine in contact with the photon thermal reservoir was proposed by Geusic et. al., who drew on the connection between the efficiency of the Carnot engine and the 3-level maser~\cite{Scovil,geusic}.  The pioneering work by Harrison and Scully~\cite{scully,harris,harris2,scully1} has demonstrated that if the incoherent light is tuned properly, the inversion of population can occur without shining laser, consequently inducing electromagnetic transparency~\cite{xiang} similar to Fano resonance~\cite{harris1}. Furthermore, photocell quantum heat engines (QHEs) were studied in the context of light harvesting complexes recently~\cite{scully1,Dorfman,Nalbach}.  In all of these studies, the blackbody radiation such as the solar light is a popular choice  for incoherent light source. However, in the presence of the metallic environment and at shorter distances from the metal, the evanescent field will be dominant~\cite{jones2}.  

Since the evanescent field near the metallic surfaces is different from the blackbody radiation due to surface polariton modes~\cite{jones2},  the non-equilibrium energy transfer in the near field region provides a new way to look at the radiation at the nanoscale different from the blackbody case~\cite{matt,rod,basu,sheila,kim,nomura}. To this end, the Forster resonant energy transfer (FRET) under the influence of the evanescent field from the metal has been studied experimentally in recent years~\cite{mivelle, wu}.  Nevertheless, it is still a topic of debate how the FRET in molecular systems is affected by the surrounding electromagnetic environment. Our previous approach~\cite{ours} to study FRET in the presence of the evanescent field was an extension of a classical model developed by Silbey and Chance to study the emission and energy transfer near the metallic surfaces~\cite{silbey2}. Statistically, the thermal near-field reservoir is similar to shared phonon reservoir~\cite{chen,silbey,david} due to the spatial correlation of the evanescent field in the vicinity of the metal, causing enhancement of the energy transfer. However, coupling of the molecular system to the evanescent field also causes greater excitation energy loss from the molecule to the metal.
 
In this paper, a quantum approach based on the polaron master equation is derived to study the population and coherence dynamics of the excitation energy in molecular systems. In the current framework, we consider the excitation energy transfer under the influence of phonon and photon baths, which play two different roles: 1. incoherent photon causes relaxation of donor-acceptor pair and the modification of dipole-dipole coupling; 2. incoherent phonon introduces the site energy broadening. The metal surface can bridge the energy transfer between donor and acceptor, albeit at the cost of losing electronic excitation energy to the metal in the form of heat, which cannot be  accounted for in a classical theory. Our quantum approach provides an important insight to the understanding of the excitation energy transfer under the influence of incoherent photon reservoir. This can provide a theoretical foundation for future studies of the quantum heat engine under the influence of the evanescent field. 

This paper consists of five sections: The dimer model in the presence of the evanescent field and phonon reservoir is discussed in Section~\ref{model}. The master equation of the dimer model is presented in Section~\ref{master}. The local density of states (LDOS) and cross spectral density of a thin metallic film are discussed in Section~\ref{ldos}. The simulation results of the population and coherence dynamics for different distances and film thicknesses can be found in Section~\ref{results}. The concluding remarks are in Section~\ref{remarks}.

\section{Dimer Model with Evanescent Field and Phonon Bath}\label{model}
For convenience, we define pseudo Pauli operators: $\sigma^{+}_j \equiv \vert j \rangle \langle 0 \vert$ and $\sigma^{-}_j \equiv \vert 0 \rangle \langle j \vert $ where $j = D, A$.  The full  Hamiltonian of the system and baths is given by: $\hat{H} = \hat{H}_S + \hat{H}_{int}  + \hat{H}_{bath}$. Here the system Hamiltonian in a single exciton manifold is:
\begin{align} 
\hat{H}_S  &\equiv \hat{H}_D + \hat{H}_A + \hat{H}_{D-A}\,,  \mbox{where} \\
\hat{H}_D &= \varepsilon_D \vert D \rangle \langle D \vert = \varepsilon_D \sigma^{+}_D \sigma^{-}_D \\
\hat{H}_A  &=  \varepsilon_A  \vert A \rangle \langle A \vert = \varepsilon_A \sigma^{+}_A \sigma^{-}_A \\ 
\hat{H}_{D-A} &=   J_{DA} (\vert D \rangle \langle A \vert  + \vert A \rangle \langle D \vert) = J_{DA} (\sigma^{+}_D \sigma^{-}_A + \sigma^{+}_A \sigma^{-}_D)\,,
\end{align}
where $J_{DA}$ is the dipole coupling between the donor dipole $\mu^{elec}_D$ and the acceptor dipole $\mu^{elec}_A$ and depends on the dyadic Green's function $\tensor{G} (\vec{r}, \vec{r}', \omega)$ of the electromagnetic environment:
\begin{align}
J_{DA}=\frac{\omega^2 \mu^{elec}_D\mu^{elec}_A}{c^2 \varepsilon_0} \vec{n}^{elec}_A \cdot \mbox{Re}[{\tensor{G}(\vec{r}_D, \vec{r}_A,  \omega) }] \cdot \vec{n}^{elec}_D\,. 
\end{align}
The coupling Hamiltonian between the donor-acceptor system and phonon and photon baths are given by:
\begin{align}
\hat{H}_{int} &= \hat{H}_{D-pn}  + \hat{H}_{A-pn}  + \hat{H}_{D-ph} +  \hat{H}_{A-ph}\,, \mbox{where}  \\
\hat{H}_{D-pn} &= \vert D \rangle \langle {D} \vert  \sum_k \hbar \lambda_{kD} (b^\dag_{k,D} + b_{k,D})  = \sigma^{+}_D \sigma^{-}_D   \sum_k \hbar \lambda_{k,D} (b^\dag_{k,D} + b_{k,D}) \\
\hat{H}_{A-pn} &= \vert {A} \rangle \langle {A} \vert \sum_l \hbar \lambda_{lA} (b^\dag_{l,A} + b_{l,A})   = \sigma^{+}_A \sigma^{-}_A \sum_l \hbar \lambda_{l,A} (b^\dag_{l,A} + b_{l,A})   \\
\hat{H}_{D-ph} &=  \vert 0 \rangle \langle {D} \vert\, \hat{\mathbb{E}}_D (\vec{r}_D, t)    +    \vert D \rangle \langle 0 \vert\, \hat{\mathbb{E}}_D^\dag (\vec{r}_D, t)  = \sigma^{+}_D\, \hat{\mathbb{E}}_D (\vec{r}_D, t)  + \sigma^{-}_D\,  \hat{\mathbb{E}}^\dag_D (\vec{r}_D, t)   \\
\hat{H}_{A-ph} &=   \vert 0 \rangle \langle {A} \vert  \, \hat{\mathbb{E}}_A (\vec{r}_A, t)   +  \vert A \rangle \langle 0\vert \, \hat{\mathbb{E}}_A^\dag(\vec{r}_A, t)  = \sigma^{+}_A\,  \hat{\mathbb{E}}_A (\vec{r}_A, t) + \sigma^{-}_A\, \hat{\mathbb{E}}^\dag_A (\vec{r}_A, t)  \,.
\end{align}
Here $\hat{\mathbb{E}}_j(\vec{r}_j, t) \equiv e^{i\omega_j t}\, \int_0^\infty d\omega\, \mu^{elec}_j\, \vec{n}^{elec}_j \cdot \vec{E}(\vec{r}_j,\omega)$, where $j = D, A$.  Note that the time dependence in operators $\hat{\mathbb{E}}$ and $\hat{\mathbb{E}}^\dag$ are explicit time-dependence in the Schrodinger picture due to dynamical dipole moment. 
The phonon and photon baths Hamiltonian are given by:
\begin{align}
&\hat{H}_{bath} = \hat{H}_{pn,D}  + \hat{H}_{pn,A}  + \hat{H}_{ph}\,, \mbox{where} \\
&\hat{H}_{pn,D} = \sum_k \hbar \omega_{k,D} b^\dag_{k,D} b_{k,D} \\
&\hat{H}_{pn,A} = \sum_l \hbar \omega_{l, A} b^\dag_{l, A} b_{l,A} \\
&\hat{H}_{ph} = \hbar \int d\vec{r} \int d\omega \, \omega \, \vec{f}^\dag(\vec{r},\omega)\,\vec{f}(\vec{r},\omega)\,.
 \end{align}
 $\hat{H}_{ph}$ is the thermal bath of photons. The electric field operator $\vec{E}(\vec{r}, \omega)$ is related to position and frequency dependent photon creation/annihilation operators $\vec{f}^\dag(\vec{r}, \omega)/\vec{f}(\vec{r}, \omega)$  by:
 \begin{align}
 \vec{E}(\vec{r},\omega)=i\frac{\omega^2}{c^2}\int d\vec{r'} \,\tensor{G}(\vec{r},\vec{r'},\omega)\sqrt{\frac{\hbar}{\epsilon_{0}}\epsilon_{I}(\vec{r'},\omega)}\vec{f}(\vec{r'},\omega)\,,
 \end{align}
where the complex relative permittivity  of the electromagnetic environment is $\epsilon (\vec{r}, \omega) = \epsilon_R(\vec{r}, \omega) + \epsilon_I (\vec{r}, \omega)$. The photon creation and annihilation operators satisfy the standard bosonic commutation relations:
\begin{align}
  &[\vec{f}(\vec{r},\omega), \vec{f}^\dag(\vec{r}',\omega')]=\delta(\vec{r}-\vec{r}')\delta(\omega-\omega')\, \\
  &[\vec{f}(\vec{r},\omega), \vec{f}(\vec{r}',\omega')]= [\vec{f}^\dag (\vec{r},\omega), \vec{f}^\dag(\vec{r}',\omega')]= 0\,.
\end{align}
In addition, the equilibrium correlations of these operators are given by:
\begin{align}
 \langle \vec{f}(\vec{r},\omega) \vec{f}^{\dagger}(\vec{r}',\omega') \rangle =  (\bar{n}(\omega) +1) \delta(\vec{r}-\vec{r}')\delta(\omega-\omega') \quad \mbox{and}  \\
  \langle \vec{f}^\dag(\vec{r},\omega) \vec{f}(\vec{r}',\omega') \rangle = \bar{n}(\omega) \delta(\vec{r}-\vec{r}')\delta(\omega-\omega')\,,
\end{align}
 where $\bar{n}(\omega)$ is the Planck function $\bar{n}(\omega) =  \frac{1}{e^{\hbar \omega/k_b T} -1}$. Using the standard relation:
 \begin{align}
 \int d\vec{s}\, \epsilon_{I}(\vec{s},\omega)\, \tensor{G}(\vec{r},\vec{s},\omega)\, \tensor{G}^{*}(\vec{s},\vec{r},\omega)= \frac{c^2}{\omega^2}\mbox{Im}[\tensor{G}(\vec{r},\vec{r}',\omega)],
 \end{align}
 we get the following correlation property for the electric field operator:
 \begin{align}
 \langle\vec{E}(\vec{r},\omega)\vec{E}^{\dagger}(\vec{r}',\omega')\rangle=\frac{\hbar \omega^2}{\epsilon_0 c^2}  [\bar{n}(\omega)+1] \mbox{Im}[\tensor{G}(\vec{r},\vec{r}',\omega)]\delta(\omega-\omega')  \\
 \langle\vec{E}^{\dagger}(\vec{r},\omega)\vec{E}(\vec{r}',\omega')\rangle=\frac{\hbar \omega^2}{\epsilon_0 c^2} \bar{n}(\omega)  \mbox{Im}[\tensor{G}(\vec{r},\vec{r}',\omega)]\delta(\omega-\omega')  \\
 \langle\vec{E}(\vec{r},\omega)\vec{E}(\vec{r}',\omega')\rangle=0\,, \quad  \langle\vec{E}^{\dagger}(\vec{r},\omega)\vec{E}^{\dagger}(\vec{r}',\omega')\rangle=0\,.
 \end{align}

\section{Master Equation}\label{master}
For completeness, we consider strong as well as weak coupling to phonon bath, while coupling to photon bath is relatively weak. In this regime, we perform polaron transformation of the Hamiltonian operator as follows: 
\begin{align}
&\hat{H}' = e^{\hat{P}}\,\hat{H}\,e^{-\hat{P}}\,, \quad \mbox{where the generating operator $\hat{P}$ is} \\
&\hat{P}  \equiv \sigma^{+}_D\,\sigma^{-}_D \hat{S}_D  + \sigma^{+}_A\,\sigma^{-}_A \hat{S}_A\,, \quad \mbox{and} \quad \hat{S}_j \equiv \sum_k \frac{\lambda_{k,j}}{\omega_{k,j}}\,(b^\dag_{k,j}-b_{k,j})\,, \quad \mbox{$j = D, A$}.
\end{align}
Furthermore, we define phonon displacement operators $\hat{B}^{\pm}_j = e^{\pm \hat{S}_j}$ where $ j = D, A $. We force the mean of the displacement operators to be zero by defining $\hat{B}^{\pm}_{0j} \equiv \hat{B}^{\pm}_j - \langle \hat{B} \rangle$, where $\langle \hat{B} \rangle =  \langle \hat{B}^{\pm}_j \rangle = e^{-\phi/2}$, where $\phi \equiv \int_0^\infty d\omega\, \frac{J_{pn}(\omega)}{\omega^2} \coth(\frac{\beta \hbar \omega}{2})$.  The polaron transformed system Hamiltonian is then given by:
\begin{align}
\hat{H}'_S =  \sum_{j=D, A} \varepsilon'_j \, \sigma^{+}_j \sigma^{-}_j  + J'_{DA} (\sigma^{+}_D\sigma^{-}_A + \sigma^{+}_A\sigma^{-}_D)\,,
\end{align}
where $\varepsilon'_j \equiv \varepsilon_j - \Delta_j$ and $J'_{DA} \equiv  J_{DA} \langle \hat{B} \rangle^2 $. The polaron shifted energy of the donor and acceptor is $\Delta_j \equiv \sum_k \frac{\lambda^2_{k,j}}{\omega_{k,j}} = \int_0^\infty d\omega\, \frac{J_{pn, j} (\omega)}{\omega}$, where $j = D, A$. The polaron transformed interaction Hamiltonian can be divided into three parts, where the first part consists of phonons only, the second part contains photons only and the third part mixes phonons with photons, respectively:
\begin{align}
&\hat{H}'_{int} = \hat{H}'_{int, 1} + \hat{H}'_{int, 2} + \hat{H}'_{int, 3}\,, \quad \mbox{where} \nonumber \\
&\hat{H}'_{int, 1} = J_{DA}\Big(\hat{B}_{DA} \sigma^{+}_D\sigma^{-}_A + \hat{B}_{AD}\sigma^{+}_A\sigma^{-}_D\Big) \\
&\hat{H}'_{int, 2} =  \langle \hat{B} \rangle \Big ( \sigma^{+}_D \hat{\mathbb{E}}(\vec{r}_D, t) + \sigma^{-}_D \hat{\mathbb{E}}^\dag(\vec{r}_D, t) + \sigma^{+}_A \hat{\mathbb{E}}(\vec{r}_A, t) + \sigma^{-}_A \hat{\mathbb{E}}^\dag(\vec{r}_A, t)\Big )\\
&\hat{H}'_{int ,3} = \hat{B}^{+}_{0D} \sigma^{+}_D \hat{\mathbb{E}}(\vec{r}_D, t) + \hat{B}^{-}_{0D} \sigma^{-}_D \hat{\mathbb{E}}^\dag(\vec{r}_D, t) + \hat{B}^{+}_{0A} \sigma^{+}_A \hat{\mathbb{E}}(\vec{r}_A, t) + \hat{B}^{-}_{0A} \sigma^{-}_A \hat{\mathbb{E}}^\dag(\vec{r}_A, t)  \nonumber \\
& \quad + \hat{B}^{+}_{0D} \sigma^{+}_A \hat{\mathbb{E}}(\vec{r}_A, t) + \hat{B}^{-}_{0D} \sigma^{-}_A \hat{\mathbb{E}}^\dag(\vec{r}_A, t) + \hat{B}^{+}_{0A} \sigma^{+}_D \hat{\mathbb{E}}(\vec{r}_D, t) + \hat{B}^{-}_{0A} \sigma^{-}_D \hat{\mathbb{E}}^\dag(\vec{r}_D, t)\,,
\end{align}
where $ \hat{B}_{ij} \equiv (\hat{B}^{+}_{0i} \hat{B}^{-}_{0j} + \langle \hat{B} \rangle \hat{B}^{+}_{0i} + \langle \hat{B} \rangle \hat{B}^{-}_{0j})$ and $j = D, A$. Note that the time dependence in operators $\hat{\mathbb{E}}$ and $\hat{\mathbb{E}}^\dag$ are explicit time-dependence in the Schrodinger picture due to dynamical dipole moment in the laboratory frame. Furthermore, the bath Hamiltonians are not affected by the polaron transformation, that is, $\hat{H}'_{bath} = \hat{H}_{bath}$.
After a polaron transformation, one can apply the  second order perturbation to compute the reduced density matrix of the system in the polaron frame:
$\hat{\rho}'_S(t) = Tr_{pn,ph}\{\hat{\rho}'(t)\} $, where $\hat{\rho}'_S(t)$ and $\hat{\rho}'(t)$ are the system and full system+bath density matrices in the polaron frame, respectively. In the interaction picture of polaron-transformed frame defined by $\hat{\tilde{O}}'(t) = \hat{U}^\dag (t)\, \hat{O}'\, \hat{U}(t)$, where $\hat{U} = \exp[-i (\hat{H}'_S+\hat{H}'_{bath}) t/\hbar]$,  the time evolution of the density matrix of the system is given by:
\begin{align}
\frac{d \hat{\tilde{\rho}}'_S}{dt} =  -\frac{1}{\hbar^2} \int_0^t d\tau\, Tr_{pn}Tr_{ph} \Big\{\Big[\hat{\tilde{H}}'_{int}(t), \Big[\hat{\tilde{H}}'_{int}(t-\tau),\hat{\tilde{\rho}}'_S(t)\hat{\rho}_{pn}\, \hat{\rho}_{ph}\Big]\Big]\Big\},
\end{align}
In the polaron frame, the master equation in the Schrodinger picture becomes:
\begin{align}
\frac{d \hat{\rho}'_S}{dt} &= \frac{1}{i \hbar} \Big[\hat{H}'_S, \hat{\rho}'_S(t) \Big] + e^{\frac{-i \hat{H}'_S t}{\hbar}} \frac{d\hat{\tilde{\rho}}'_S}{dt} e^{\frac{i \hat{H}'_S t}{\hbar}} \nonumber \\
\label{MasterEq}
&= \frac{1}{i\hbar} \Big[\hat{H}'_S, \hat{\rho}'_S(t) \Big]  - \frac{1}{\hbar^2}\,\int_0^t d\tau\, Tr_{pn}Tr_{ph} \Big\{\Big[\hat{\tilde{H}}'_{int}(0), \Big[\hat{\tilde{H}}'_{int}(-\tau),\hat{\rho}'_S(t)\hat{\rho}_{pn}\, \hat{\rho}_{ph}\Big]\Big]\Big\} 
\end{align}
Here the time-dependence of an operator $\hat{\tilde{O}}(t)$ is defined by $\hat{\tilde{O}}(t) = \hat{U}^\dag(t)\,\hat{O}\, \hat{U}(t)$, where $\hat{U} = \exp[-i (\hat{H}'_S+\hat{H}'_{bath}) t/\hbar]$. The detailed expansion of the terms in Eq.~\ref{MasterEq} are provided in section~\ref{apndx}. Below, we solve the master equation numerically in polaron frame. The observables in the lab frame $\langle \hat{O}\rangle(t)$ are obtained from the density matrix computed in the polaron frame $\hat{\rho}'_S(t)$ in the following way:
\begin{align}
\langle \hat{O}\rangle (t) = Tr_S Tr_{pn} [\hat{\rho}_S (t) \,\hat{O}] =  Tr_S Tr_{pn}[e^{-\hat{P}}\, \hat{\rho}'_S (t) \,e^{\hat{P}} \,\hat{O}] = Tr_S Tr_{pn} [e^{\hat{P}}\,\hat{O}\, e^{-\hat{P}}\,\hat{\rho}'_S (t) ] 
\end{align}
The population of the donor/acceptor states (or diagonal elements) remain invariant under the polaron transformation. However, the coherence or the off-diagonal elements are affected by the polaron transformation. For instance, if $\hat{O} = \sigma^{+}_D\,\sigma^{-}_A$, then we have
\begin{align}
\langle \sigma^{+}_D\,\sigma^{-}_A \rangle (t) &= Tr_S Tr_{pn}[e^{\hat{P}} \, \sigma^{+}_D\,\sigma^{-}_A\, e^{\hat{P}}\, \hat{\rho}'_S(t)] \nonumber \\
&= Tr_S Tr_{pn}[\sigma^{+}_D\,\sigma^{-}_A\,\hat{B}^{+}_D\,\hat{B}^{-}_A\, \hat{\rho}'_S(t)] = \langle \hat{B} \rangle^2 Tr_S[\sigma^{+}_D\, \sigma^{-}_A\,\hat{\rho}'_S(t)]
\end{align}

\section{Thin Film and Evanescent Field}\label{ldos}

The statistical properties of the evanescent field are determined by the dyadic Green's tensor. For thin films, it is possible to compute the dyadic Green's function analytically. Such geometries are also useful if we limit ourselves to the situation where the separation of the donor-acceptor system from the metal surface is smaller than the radius of curvature of the surface so that the surface can be assumed to be flat. For a source point at $\vec{r}=(x, y, z)$, a field point at $\vec{r}' = (x', y', z')$ and defining a two dimensional vector $\vec{\rho} = (x-x', y-y')$, the scattering Green's function for a half space or a thin film geometry with the material permittivity $\varepsilon_2$ in both local and non-local limits, and the surrounding permittivity $\varepsilon_1$ is given by:
\begin{align}
\label{Dxx_Dzz}
&\tensor{G}_{xx}(\vec{r}, \vec{r}', \omega) =\frac{i c^2}{8 \pi ^2\omega^2} \int_{0}^{2 \pi} d\theta \int_{0}^{\infty}\frac{pdp}{\varepsilon_1 q_1}\,e^{iq_1(z+z') + i\vec{p}\cdot\vec{\rho}}  \nonumber \\
&\times \Big[\frac{\omega^2}{c^2}r_{s}(p) \sin^2\theta -q_1^2r_{p}(p) \cos^2\theta\Big] \,,\\
&\tensor{G}_{zz}(\vec{r}, \vec{r}', \omega)=\frac{i c^2}{8\pi^2\omega^2} \int_{0}^{2\pi} d\theta \int_{0}^{\infty}\frac{p^3}{\varepsilon_1 q_1}dp \nonumber \\
&\times e^{iq_1(z+z') + i \vec{p}\cdot\vec{\rho}} \,\, r_{p}(p) \,,
\end{align}
where $p$ is the transverse  and $q_1$ is the z-component of the wave vector, with $q_1=\sqrt{\omega^2/c^2\varepsilon_1-p^2}$. All other components of the dyadic Green's function can be computed from these components. For a thin film geometry of thickness $a$ and the local permittivity $\varepsilon_2$ surrounded by another medium of permittivity $\varepsilon_1$ on both sides of the thin film, one can also derive the Fresnel reflection coefficients analytically~\cite{}. The results are:
\begin{subequations}
\begin{align}
r_s(p) = \frac{q_1^2 - q_2^2}{q_2^2 +q_1^2 + 2i q_1 q_2 \cot(q_2 a)} \\
r_p(p) = \frac{\varepsilon_2^2 q_1^2 - \varepsilon_1^2 q_2^2}{\varepsilon_1^2q_2^2 + \varepsilon_2^2 q_1^2 + 2i \varepsilon_1 \varepsilon_2 q_1 q_2 \cot(q_2 a)}
\end{align}
\end{subequations}
where $q_2 = \sqrt{\omega^2/c^2\varepsilon_2 - p^2} $.

\begin{figure}
 \includegraphics[width=0.9\columnwidth]{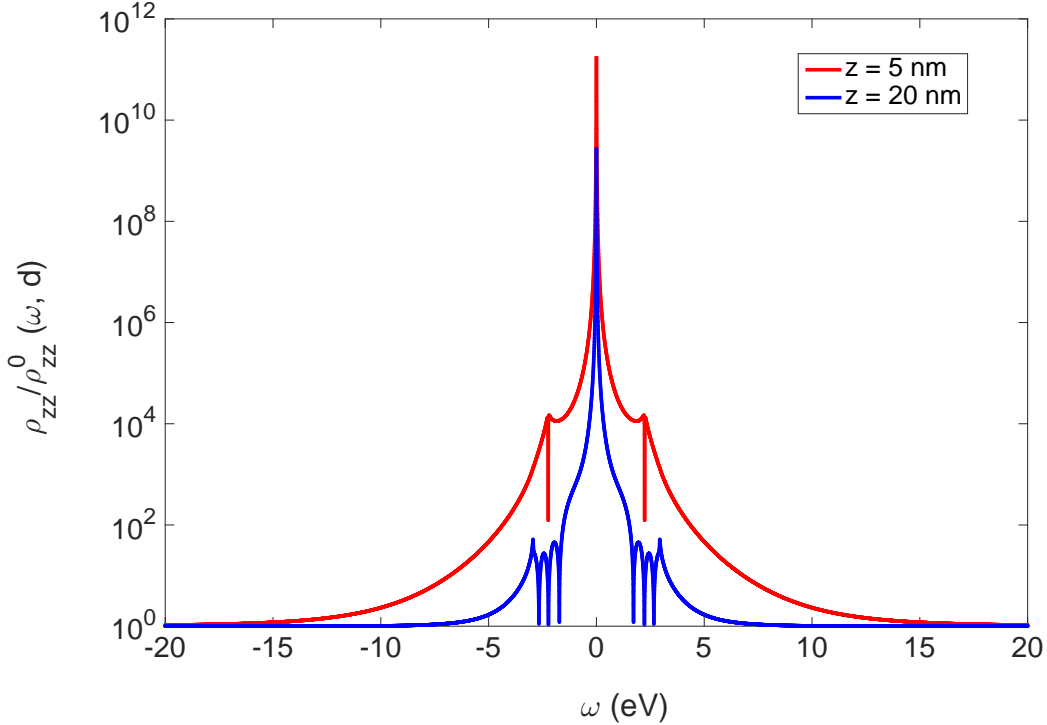}
\caption{The ration of the z-component of the electric  local density of states  in the vicinity of a silver thin film to free space. Thin film is of thickness $a = 10$ nm and at two different distances from the surface of the thin film: $z = 5$ nm (solid red) and $z=20$ nm (solid blue).}
\label{LDOS}
\end{figure}

\begin{figure}
 \includegraphics[width=0.9\columnwidth]{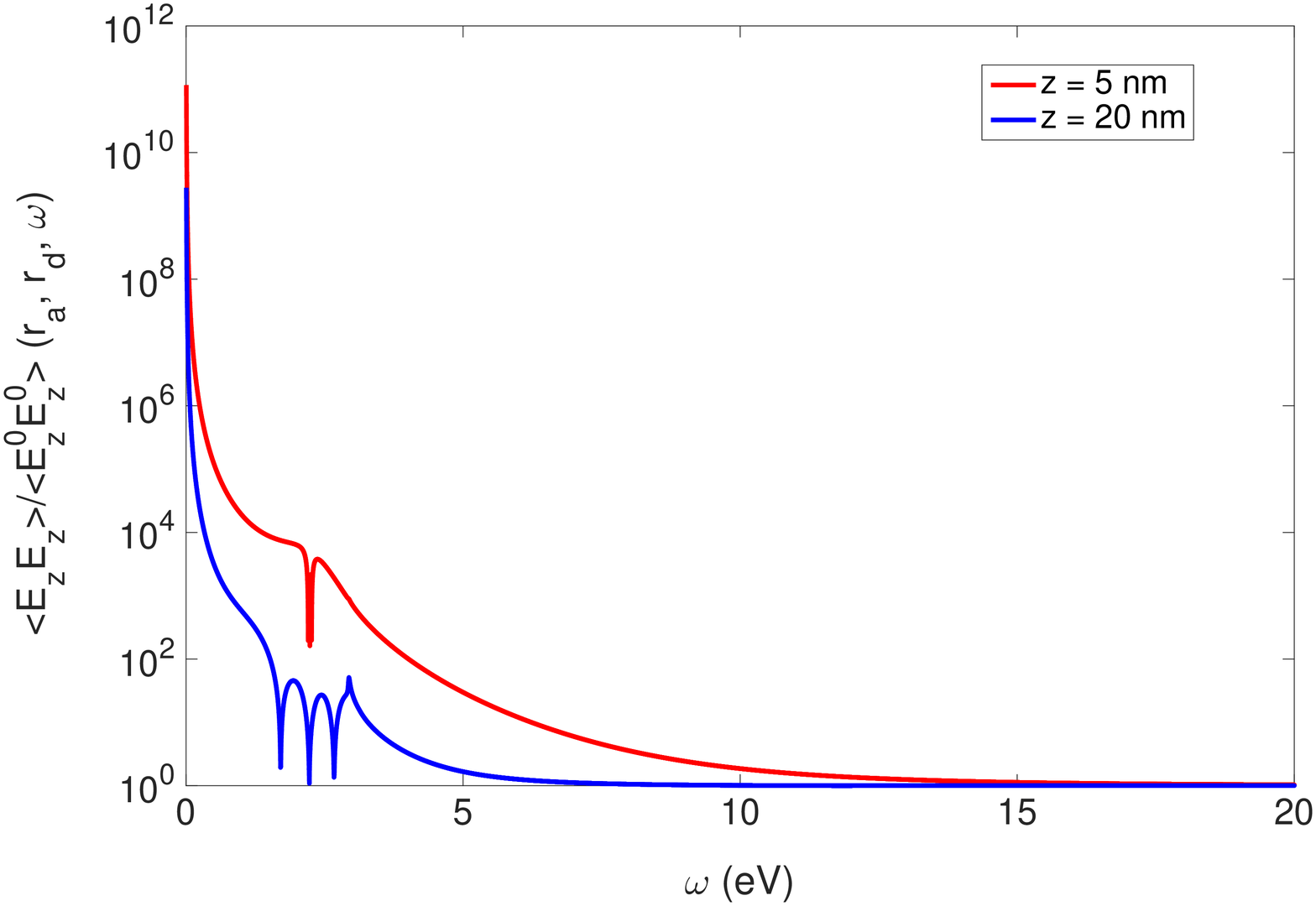}
\caption{The ration of the z-component of the electric cross spectral density  in the vicinity of a silver thin film to free space. Thin film is of thickness $a = 10$ nm and at two different distances from the surface of the thin film: $z= 5$ nm (solid red) and $z=20$ nm (solid blue). The donor acceptor separation distance $x=4$ nm.}
\label{CSD}
\end{figure}

There are two key factors in the thin film model, 1. Distance from the surface of the thin film (geometrical factor) 2. dielectric constant (material property). When these two factors change, the local density of state (LDOS) and cross correlation of the electric part of the photon field will change as well. 
The LDOS (electric part only) can be defined as \cite{joulain}
\begin{equation}
\rho(\vec{r},\omega) = \frac{\omega}{\pi c^2} \text{Im} \Big[ \text{Tr} [\tensor{D}^E(\vec{r},\vec{r}, \omega)]\Big]
\end{equation}
The cross correlation of the electric field at donor and acceptor locations is given by:
\begin{equation}
\mathcal{E}(\vec{r},\vec{r}',\omega) = \langle E^*_i(\vec{r},\omega) E_j(\vec{r}',\omega) \rangle = \frac{\hbar \omega^2}{\epsilon_0 c^2} \text{Im}\Big[\tensor{D}_{ij}^E(\vec{r},\vec{r}',\omega)\Big].
\end{equation}
It is clear from the above expressions that the LDOS and the cross spectral density are affected by the geometrical parameter, namely, the distance from and thickness of the  film. In Fig.~\ref{LDOS}, we plot the z-component of LDOS  of the electric field for a thin film of thickness $a = 10$ nm and at distances of $z = 5$ nm and $20$ nm along z-axis from the surface of the thin film. Furthermore, we plot an electric cross spectral density $\mathcal{E}_{ij}(r_A,r_D,\omega)$  for the same set of parameters  in Fig.~\ref{CSD}.

\section{Dynamics of Excitation Transfer}\label{results}
In this section we discuss the dynamics of exciton transfer in the presence of thin metallic film. In the vicinity of a metallic thin film,  the LDOS of the evanescent field is greatly modified. This modification leads to enhanced relaxation rate of a molecule in the vicinity of a thin film. However, the presence of metal also greatly modifies the cross spectral density which is responsible for the exciton transfer from donor to acceptor.  Since the LDOS and cross spectral densities are determined by the geometrical shape of the electromagnetic environment, we investigate the energy transfer quantum dynamics by varying different geometrical parameter, namely, the distance between the donor-acceptor pair and a thin film, separation distance between the donor and acceptor and the thickness of a thin film. Due to the spatial correlation of the evanescent field, the donor and acceptor can transfer energy through the evanescent modes. 

We consider a symmetric dimer model with donor and acceptor dipole moments  $\mu^{\text{elec}}_D = \mu^{\text{elec}}_A = 1$ Debye, pointing along the z-axis, without any loss of generality. The distance between the donor and acceptor is $d = 2$ nm unless otherwise. We consider the donor and acceptor wavelength of $\lambda = 630$ nm and the ambient temperature $T = 300$ K. We model the permittivity of a metallic thin film using a Drude model with a plasmon frequency $\omega_p = 4.6\times10^{15}$ rad/s and an electron scattering rate $\nu = 3.4\times10^{13} $ rad/s.

\subsection{Population Dynamics}
\begin{figure}
 \includegraphics[width=0.9\columnwidth]{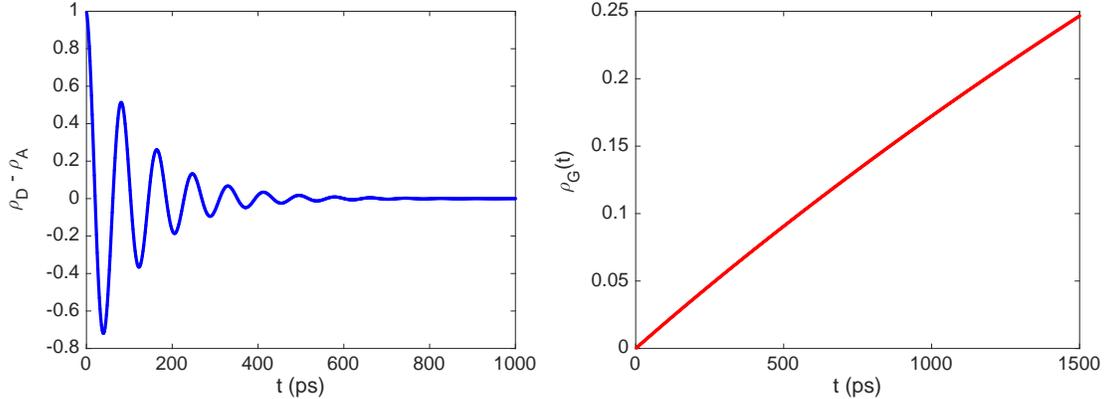}
\caption{The dynamics of population difference between the donor $D$ and acceptor $A$ states in the vicinity of a metallic thin film. The donor and acceptor are located at a distance $z = 10$ nm from a metallic thin film of thickness $a = 10$ nm and are separated from each other by a distance $d = 2$ nm. The plot on the right panel shows the ground state population.}
\label{f1}
\end{figure}

In Fig.~\ref{f1}, we plot the dynamics of the population difference between the donor $D$ and acceptor $A$ states in the vicinity of a metallic thin film. Here we consider weak exciton-phonon interaction characterized by a coupling parameter $\Gamma = 0.1$. We fix the donor and acceptor distance $d = 2$ nm and the distance of donor-acceptor pair from the metallic thin film is $z = 10$ nm. For this choice of parameters, we observe oscillation in donor and acceptor population. However, we also find that the ground state is populated, which results from the absorption of the energy by the metal. At this distance from the metal and weak exciton-phonon coupling, exciton transfer is not greatly affected by the presence of the metal. 

\begin{figure}
 \includegraphics[width=0.9\columnwidth]{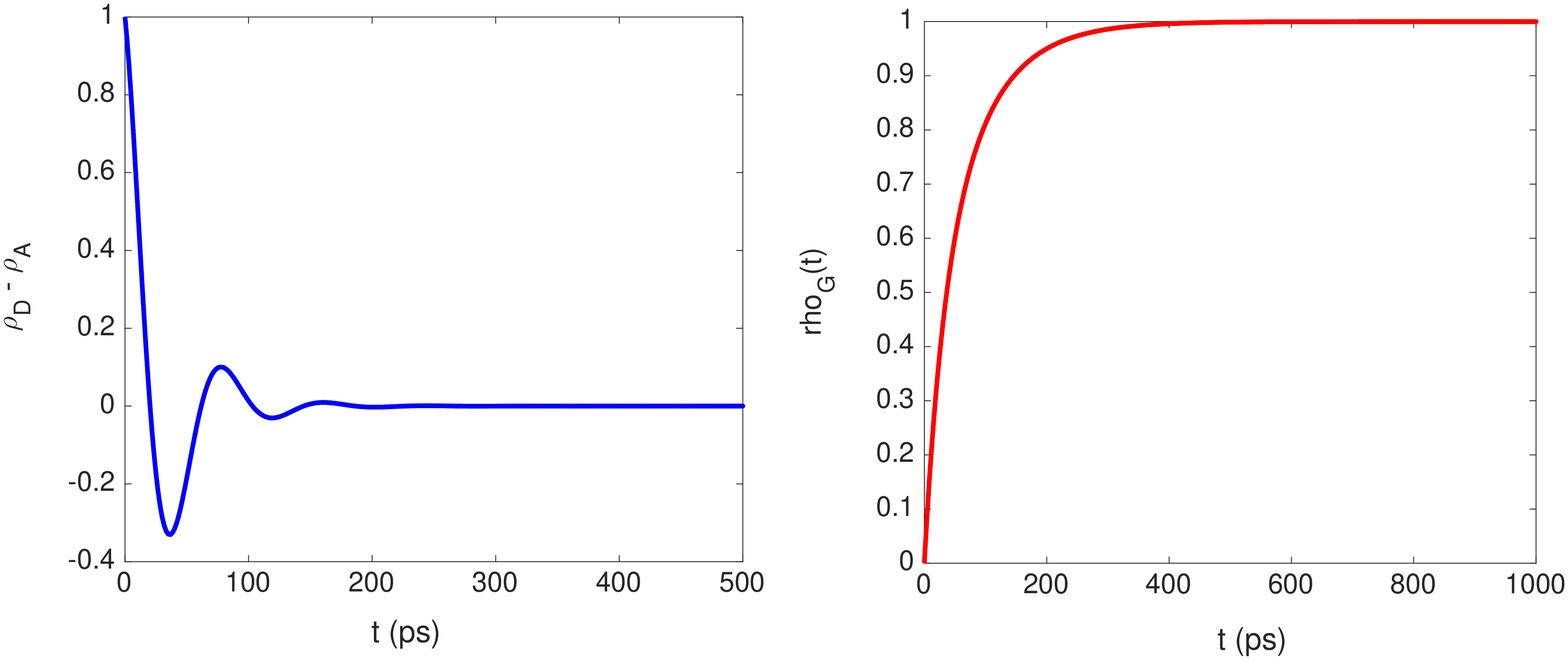}
\caption{The dynamics of population difference between the donor $D$ and acceptor $A$ states in the vicinity of a metallic thin film. The donor and acceptor are located at a distance $z = 2$ nm from a metallic thin film of thickness $a = 10$ nm and are separated from each other by a distance $d = 2$ nm. The plot on the right panel shows the ground state population.}
\label{f2}
\end{figure}

Next, we consider a smaller distance from the metal. In Fig.~\ref{f2}, we plot the population difference between the donor and acceptor located at a distance $z = 2$ nm from a thin metallic film. At a short distance from the metal, exciton transfer from donor to acceptor is much faster compared to larger distances. However, at such a short distance, the energy is also absorbed by the metal at a much faster rate, as indicated by the ground state population shown in the right panel of Fig.~\ref{f2}. 

\begin{figure}
 \includegraphics[width=0.9\columnwidth]{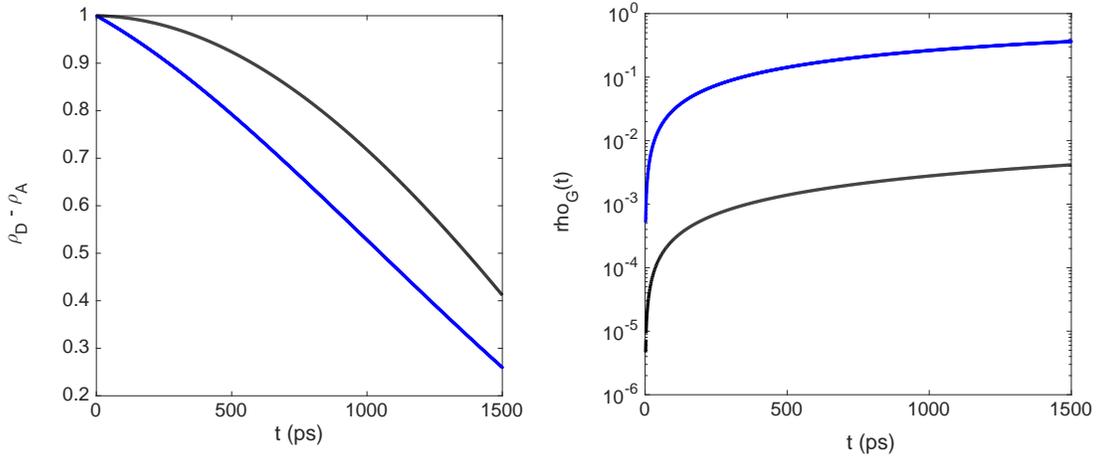}
\caption{The dynamics of population difference between the donor $D$ and acceptor $A$ states in the vicinity of a metallic thin film in the presence of strong exciton-phonon coupling. The donor and acceptor are located at a distance $z = 2$ nm (solid blue) and $z = 10$ nm (dashed black) from a metallic thin film of thickness $a = 10$ nm and are separated from each other by a distance $d = 2$ nm. The plot on the right panel shows the ground state population.}
\label{f3}
\end{figure}

We now consider a strong coupling of donor-acceptor pair to phonon modes, characterized by exciton-phonon coupling parameter $\Gamma =1$. In Fig.~\ref{f3}, we plot the population difference of donor-acceptor pair in the strong coupling regime. In this regime, the exciton dynamics does not reveal any oscillations and the energy transfer  process is dominated by the incoherent transfer process. We plot the dynamics at two different distances from the metal: $z = 2$ nm (solid blue) and $z=10$ nm (dashed black).  These plots reveal that exciton transfer is faster at a smaller distance than longer distance from thin film, even in the regime of strong coupling to phonons.

\begin{figure}
 \includegraphics[width=0.9\columnwidth]{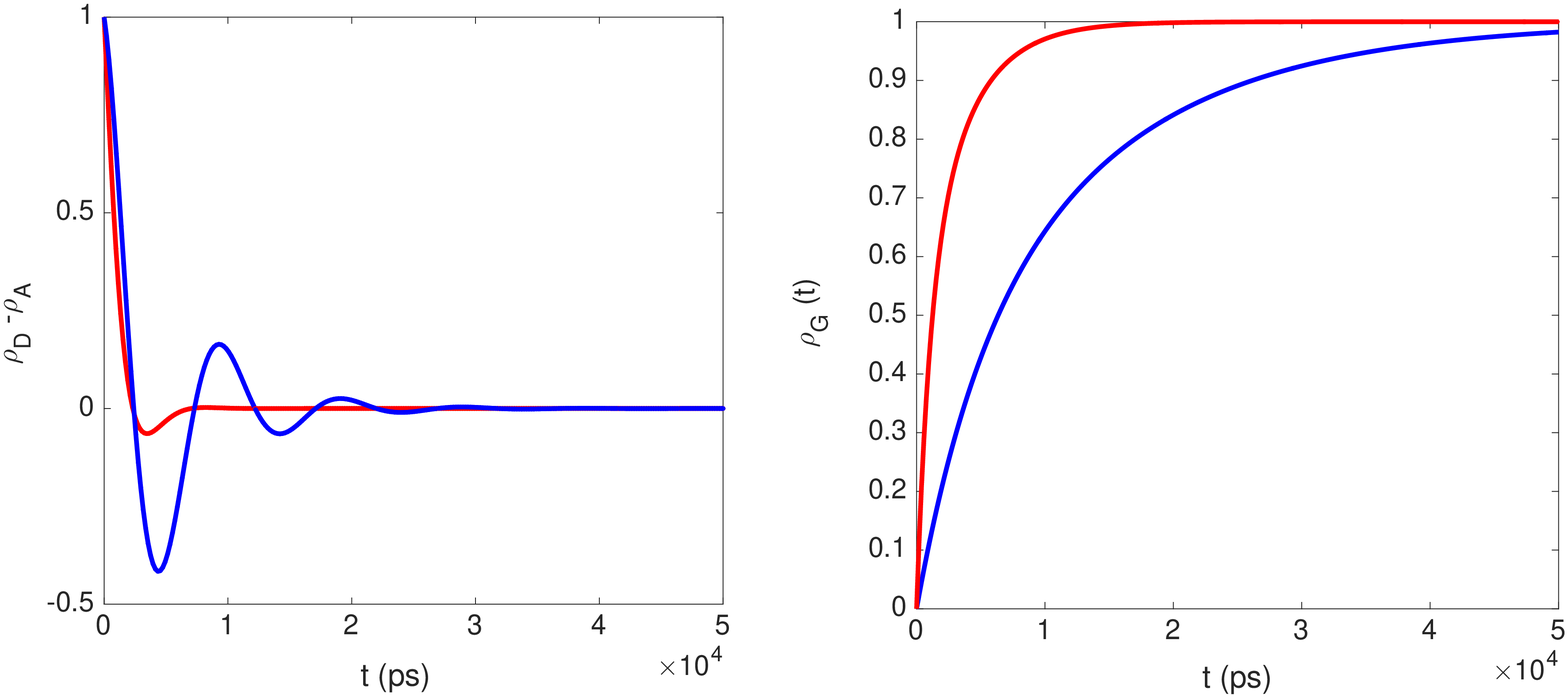}
\caption{The dynamics of population difference between the donor $D$ and acceptor $A$ states in the vicinity of a metallic thin film of various thickness: $a=5$ nm (solid blue) and $a=50$ nm (solid red). The donor and acceptor are located at a distance $z = 10$ nm  from a metallic thin film and are separated from each other by a distance $d = 10$ nm. The plot on the right panel shows the ground state population.}
\label{f4}
\end{figure}

In the following, we discuss the effect of the thickness of a thin film on the exciton dynamics. In Fig.~\ref{f4}, we plot the donor acceptor population difference for two different thickness of a thin film. On the left panel we also plot the ground state population. We place the donor and acceptor at a distance $d = 10$ nm from each other and at a distance $z=10$ nm from the surface of the thin film. For a film thickness of $a = 5$ nm, we find that the population dynamics exhibit oscillatory behavior. However, for thin film of thickness $a=50$ nm, oscillation is strongly suppressed. The rapid increase in the ground state population in the vicinity of a thick metallic film, indicates that exciton transfer is relatively more efficient for a thin film whose thickness is smaller than the separation distance between the donor-acceptor pair.

\subsection{Coherent Dynamics}
\begin{figure}
 \includegraphics[width=0.9\columnwidth]{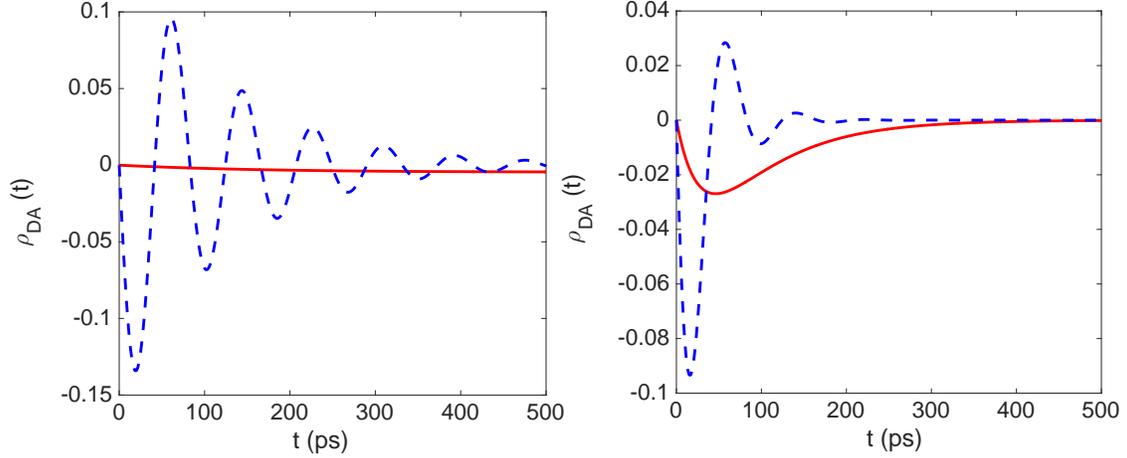}
\caption{The coherence between donor-acceptor pair in the vicinity of a thin film and weak electron phonon coupling. The real (solid red) and imaginary (dashed blue) parts of the off-diagonal elements of the density matrix of donor-acceptor pair. The donor and acceptor are located at a distance $z = 10$ nm (left panel) and $z =2$ nm (right panel) from a metallic thin film of thickness $a = 10$ nm and are separated from each other by a distance $d = 2$ nm.}
\label{f5}
\end{figure}

In this section,  we discuss the coherences of donor and acceptor pair characterized by the off-diagonal matrix elements $\rho_{DA}$ at different distances from the metallic thin film. In Fig.\ref{f5}, we plot the off-diagonal elements of the donor-acceptor density matrix at different distances from the metal. We find that at distances comparable to film thickness, the coherence decay is much slower than at distances smaller than film thickness. While the energy transfer is faster when the donor-acceptor pair is closer to the surface of the metal, the phase coherence decays rapidly at smaller distances from the metal. The loss of phase coherence at smaller distances is due to energy loss by the donor-acceptor system to the metal.

\section{Conclusions}\label{remarks}
In this paper, a quantum approach to excitation energy transfer in the vicinity of a metal has been proposed. Here, we have considered coupling of an exciton to phonon as well as photon baths. A small polaron transformation has been applied and the polaron master equation has been derived including the electromagnetic effect due to the surrounding metallic surface. We have constructed the master equation in the polaron frame and provided the transformation scheme from the polaron to the lab frame. Due to the polaron transformation, the dipole moment can shift so that  the intermediate region still needs a more special treatment. In this paper, we have only considered the weak and strong coupling regimes. The intermediate coupling of phonon with system can introduce the steady state distribution beyond the Fermi's golden rule. Different from the emission and energy relaxation processes, the energy transfer process is more efficient if dissipation through the non-radiative channel is small. Furthermore, our calculations have demonstrated that there is competition between dimer separation and thickness of thin film which will subsequently affect the population and coherence dynamics. We have found that when the thickness of a metallic film is smaller than the donor-acceptor separation, then the population oscillation can last longer.

Our calculations have demonstrated that there is a competition between two energy transfer processes, namely, the energy transfer from donor to acceptor and the energy transfer (or energy dissipation) from donor-acceptor pair to the metal. Our model has indicated that if the energy transfer rate is much faster than the energy dissipation rate to the metal, the efficiency of the energy transfer can be relatively higher. To this end, our calculation delivers an important message: metallic surfaces cannot enhance transfer rate and transfer efficiency simultaneously due to the lossy metallic environment. Finding alternative surfaces, such as meta-material, which can reduce the loss of excitation energy, can potentially enhance both transfer rate and efficiency. This will be discussed in our future publications. However, our calculations have not considered how fast the excitation energy is being used, which is often at the rate of charge separation. When the excitation energy is consumed by the acceptor at a rate faster than the dissipation rate to the metal,  then the efficiency may also be enhanced even in the presence of lossy metallic environment.

While quantum heat engine has been used to study the efficiency of the system of the excitation energy transfer in a molecular network, such as, FMO\cite{scully1}, our model can serve as a foundation to study how the evanescent field, other than the blackbody radiation (propagating far field), can be used to boost the efficiency of the system of the excitation energy transfer in a molecular network. Many parameters, such as the energy network (funnel effect), geometry (distance, thickness and curvature) and material (dielectric), etc. can be tuned to increase the transfer rate and efficiency, which demonstrates versatility of the model developed in this paper.

\section{Appendix}\label{apndx}
We split the double commutators in Eq.~\ref{MasterEq} into several non-zero terms. This is possible since the average value of the product of any two terms of the  interaction Hamiltonian go to zero and hence each term can be treated separately: 
\begin{align}
&Tr_{pn}Tr_{ph} \Big\{\Big[\hat{\tilde{H}}'_{int}(0), \Big[\hat{\tilde{H}}'_{int}(-\tau),\hat{\tilde{\rho}}'_S(t)\hat{\rho}_{pn}\, \hat{\rho}_{ph}\Big]\Big]\Big\} = (I) + (II) + (III)  \\
\label{h11}
&(I) = \Big\langle \Big[\hat{\tilde{H}}'_{int, 1}(0)\,, \Big[\hat{\tilde{H}}'_{int, 1}(-\tau)\,, \hat{{\rho}}'_S(t)\Big]\Big] \Big\rangle_{pn} \\
\label{h22}
&\quad (II) =  \Big\langle \Big [\hat{\tilde{H}}'_{int, 2}(0) \,, \Big[\hat{\tilde{H}}'_{int, 2}(-\tau) \,, \hat{{\rho}}'_S(t) \Big] \Big]\Big\rangle_{ph}\\
\label{h33}
&\quad (III) = \Big\langle \Big[ \hat{\tilde{H}}'_{int, 3}(0) \,, \Big[\hat{\tilde{H}}'_{int, 3}(-\tau) \,, \hat{{\rho}}'_S(t) \Big]\Big]\Big\rangle_{pn, ph} 
\end{align}
where 
\begin{align}
&\hat{\tilde{H}}'_{int, 1}(t) = J_{DA}\Big( \hat{\tilde{B}}_{DA}(t) \tilde{\sigma}^{+}_D(t) \tilde{\sigma}^{-}_A (t)+ \hat{\tilde{B}}_{AD}(t) \tilde{\sigma}^{+}_A (t) \tilde{\sigma}^{-}_D (t)\Big) \\
&\hat{\tilde{H}}'_{int, 2}(t) =  \langle \hat{B} \rangle \Big ( \tilde{\sigma}^{+}_D(t) \hat{\tilde{\mathbb{E}}}(\vec{r}_D, t) + \tilde{\sigma}^{-}_D(t) \hat{\tilde{\mathbb{E}}}^\dag(\vec{r}_D, t) + \tilde{\sigma}^{+}_A(t) \hat{\tilde{\mathbb{E}}}(\vec{r}_A, t) + \tilde{\sigma}^{-}_A(t) \hat{\tilde{\mathbb{E}}}^\dag(\vec{r}_A, t)\Big )  \\
&\hat{\tilde{H}}'_{int, 3}(t) =\hat{\tilde{B}}^{+}_{0D}(t) \tilde{\sigma}^{+}_D(t) \hat{\tilde{\mathbb{E}}}(\vec{r}_D, t) + \hat{\tilde{B}}^{-}_{0D}(t) \tilde{\sigma}^{-}_D(t) \hat{\tilde{\mathbb{E}}}^\dag(\vec{r}_D, t) + \hat{\tilde{B}}^{+}_{0A}(t) \tilde{\sigma}^{+}_A(t) \hat{\tilde{\mathbb{E}}}(\vec{r}_A, t) + \hat{\tilde{B}}^{-}_{0A}(t) \tilde{\sigma}^{-}_A(t) \hat{\tilde{\mathbb{E}}}^\dag(\vec{r}_A, t)  \nonumber \\
& \quad + \hat{\tilde{B}}^{+}_{0D}(t) \tilde{\sigma}^{+}_A(t) \hat{\tilde{\mathbb{E}}}(\vec{r}_A, t) + \hat{\tilde{B}}^{-}_{0D}(t) \tilde{\sigma}^{-}_A(t) \hat{\tilde{\mathbb{E}}}^\dag(\vec{r}_A, t) + \hat{\tilde{B}}^{+}_{0A}(t) \tilde{\sigma}^{+}_D(t) \hat{\tilde{\mathbb{E}}}(\vec{r}_D, t) + \hat{\tilde{B}}^{-}_{0A}(t) \tilde{\sigma}^{-}_D(t) \hat{\tilde{\mathbb{E}}}^\dag(\vec{r}_D, t)\,.
\end{align}
We expand Eq.~\ref{h11} and simplify the terms in the expansion to get the following expression:
\begin{align}
(I) &= J^2_{DA} \Bigg \{ \Big [ \beta_{DAAD}(0, -\tau) \, \sigma^{+}_D \sigma^{-}_A \tilde{\sigma}^{+}_A(-\tau) \tilde{\sigma}^{-}_D(-\tau)  + \beta_{ADDA}(0, -\tau)\,  \sigma^{+}_A \sigma^{-}_D \tilde{\sigma}^{+}_D(-\tau) \tilde{\sigma}^{-}_A(-\tau) \Big] \,\hat{\rho}'_S(t) \nonumber \\
&- \Big [\beta_{ADDA}(-\tau, 0)\,  \sigma^{+}_D \sigma^{-}_A\, \hat{{\rho}}'_S(t)\, \tilde{\sigma}^{+}_A(-\tau) \tilde{\sigma}^{-}_D(-\tau) + \beta_{DAAD}(-\tau, 0)\,  \sigma^{+}_A \sigma^{-}_D \, \hat{{\rho}}'_S(t)\, \tilde{\sigma}^{+}_D(-\tau) \tilde{\sigma}^{-}_A(-\tau) \Big]  \nonumber \\
&- \Big [\beta_{ADDA}(0, -\tau)\,  \tilde{\sigma}^{+}_D(-\tau) \tilde{\sigma}^{-}_A(-\tau) \, \hat{{\rho}}'_S(t)\, \sigma^{+}_A \sigma^{-}_D + \beta_{DAAD}(0, -\tau) \,  \tilde{\sigma}^{+}_A(-\tau) \tilde{\sigma}^{-}_D(-\tau) \, \hat{\rho}'_S(t)\, \sigma^{+}_D \sigma^{-}_A\Big]  \nonumber \\
&+ \hat{\rho}'_S(t)\, \Big [\beta_{DAAD}(-\tau, 0) \,  \tilde{\sigma}^{+}_D(-\tau) \tilde{\sigma}^{-}_A(-\tau) \sigma^{+}_A \sigma^{-}_D + \beta_{ADDA}(-\tau, 0) \,  \tilde{\sigma}^{+}_A(-\tau)\tilde{\sigma}^{-}_D(-\tau) \sigma^{+}_D \sigma^{-}_A \Big]  \Bigg \}
\end{align}

The phonon correlation functions $\beta_{ijkl}(t, t')$ [note that $\beta_{ijkl}(t', t) = \beta^*_{ijkl}(t, t')$]  are given by:
\begin{align}
&\beta_{DADA}(t, t') \equiv \langle \hat{\tilde{B}}_{DA}(t)\,\hat{\tilde{B}}_{DA}(t')\rangle =  \langle \hat{B} \rangle^4 (e^{-2\phi(t-t')} -1) \\
&\beta_{DAAD}(t, t') \equiv \langle \hat{\tilde{B}}_{DA}(t)\,\hat{\tilde{B}}_{AD}(t')\rangle =  \langle \hat{B} \rangle^4 (e^{2\phi(t-t')} -1)\\
&\beta_{ADDA}(t, t') \equiv \langle \hat{\tilde{B}}_{AD}(t)\,\hat{\tilde{B}}_{DA}(t')\rangle =  \langle \hat{B} \rangle^4 (e^{2\phi(t-t')} -1) \\
&\beta_{ADAD}(t, t') \equiv \langle \hat{\tilde{B}}_{AD}(t)\,\hat{\tilde{B}}_{AD}(t')\rangle =  \langle \hat{B} \rangle^4 (e^{-2\phi(t-t')} -1)\,,
\end{align}
where we have used the following correlation functions of the mean-zero displacement operators:
\begin{align}
\langle \hat{\tilde{B}}^{\pm}_{0i}(t)\, \hat{\tilde{B}}^{\pm}_{0j}(t') \rangle =  
\begin{cases}
\langle \hat{B} \rangle^2 \, (e^{-\phi(t-t')} -1)\,, \quad \mbox{if $i = j$} \\
0\,, \quad \mbox{if $i \neq j$, where $i, j = D, A$.}
\end{cases}
\end{align}
\begin{align}
\langle \hat{\tilde{B}}^{\pm}_{0i}(t)\, \hat{\tilde{B}}^{\mp}_{0j}(t') \rangle =  
\begin{cases}
\langle \hat{B} \rangle^2 \, (e^{\phi(t-t')} -1)\,, \quad \mbox{if $i = j$} \\
0\,, \quad \mbox{if $i \neq j$, where $i, j = D, A$.}
\end{cases}
\end{align}
and the function $\phi(t-t')$ is given by:
\begin{align}
\phi(t-t') &=  \sum_k \frac{\lambda^2_k}{\omega^2_k} \Big[ \bar{n}(\omega_k) e^{i \omega_k (t-t')} + \{1 + \bar{n}(\omega_k) \} e^{-i \omega_k (t-t')} \Big ] \nonumber \\
& = \int_0^\infty d\omega \, \frac{J_{pn}(\omega)}{\omega^2} \Big[\coth(\beta \hbar \omega/2) \cos(\omega (t-t')) - i \sin(\omega (t-t')) \Big]\,.
\end{align}
Next, we expand Eq.~\ref{h22} and obtain the following expression:
\begin{align}
(II) &= \langle \hat{B} \rangle^2 \Bigg\{ \Big [\alpha_{DD}(0, -\tau) \sigma^{+}_D \tilde{\sigma}^{-}_D(-\tau)  + \alpha_{DA}(0, -\tau) \sigma^{+}_D \tilde{\sigma}^{-}_A(-\tau) \nonumber \\
& + \bar{\alpha}_{DD} (0, -\tau)\sigma^{-}_D \tilde{\sigma}^{+}_D(-\tau)  + \bar{\alpha}_{DA}(0, -\tau) \sigma^{-}_D \tilde{\sigma}^{+}_A(-\tau) \nonumber \\
& + \alpha_{AD}(0, -\tau) \sigma^{+}_A \tilde{\sigma}^{-}_D(-\tau)  + \alpha_{AA}(0, -\tau) \sigma^{+}_A \tilde{\sigma}^{-}_A(-\tau) \nonumber \\
& + \bar{\alpha}_{AD}(0, -\tau) \sigma^{-}_A \tilde{\sigma}^{+}_D(-\tau)  + \bar{\alpha}_{AA}(0, -\tau) \sigma^{-}_A \tilde{\sigma}^{+}_A(-\tau)\Big] \, \hat{{\rho}}'_S (t) \nonumber \\
&- \Big [\bar{\alpha}_{DD}(-\tau, 0) \sigma^{+}_D \, \hat{{\rho}}'_S (t) \, \tilde{\sigma}^{-}_D(-\tau)  + \bar{\alpha}_{DA}(-\tau, 0) \sigma^{+}_D \, \hat{{\rho}}'_S (t) \, \tilde{\sigma}^{-}_A(-\tau) \nonumber \\
&+ \alpha_{DD}(-\tau, 0)\sigma^{-}_D \, \hat{{\rho}}'_S (t) \,  \tilde{\sigma}^{+}_D(-\tau)  + \alpha_{DA}(-\tau, 0) \sigma^{-}_D \, \hat{{\rho}}'_S (t) \,  \tilde{\sigma}^{+}_A(-\tau) \nonumber \\
& +  \bar{\alpha}_{AD}(-\tau, 0) \sigma^{+}_A \, \hat{{\rho}}'_S (t) \,  \tilde{\sigma}^{-}_D(-\tau)  + \bar{\alpha}_{AA}(-\tau, 0) \sigma^{+}_A \, \hat{{\rho}}'_S (t) \,  \tilde{\sigma}^{-}_A (-\tau) \nonumber \\
&+  \alpha_{AD}(-\tau, 0) \sigma^{-}_A \, \hat{{\rho}}'_S (t) \,  \tilde{\sigma}^{+}_D(-\tau)  +  \alpha_{AA}(-\tau, 0)  \sigma^{-}_A \, \hat{{\rho}}'_S (t) \, \tilde{\sigma}^{+}_A(-\tau) \Big] \nonumber \\
&-\Big [\bar{\alpha}_{DD}(0, -\tau)\tilde{\sigma}^{+}_D(-\tau) \, \hat{{\rho}}'_S (t) \,  \sigma^{-}_D  + \bar{\alpha}_{DA}(0, -\tau)\tilde{\sigma}^{+}_D(-\tau) \, \hat{{\rho}}'_S (t) \,  \sigma^{-}_A  \nonumber \\
& + \alpha_{DD}(0, -\tau) \tilde{\sigma}^{-}_D(-\tau)  \, \hat{{\rho}}'_S (t) \, \sigma^{+}_D  + \alpha_{DA}(0, -\tau) \tilde{\sigma}^{-}_D(-\tau) \, \hat{{\rho}}'_S (t) \,  \sigma^{+}_A \nonumber \\
& + \bar{\alpha}_{AD}(0, -\tau)\tilde{\sigma}^{+}_A(-\tau) \, \hat{{\rho}}'_S (t) \,  \sigma^{-}_D  +  \bar{\alpha}_{AA} (0, -\tau)\sigma^{+}_A(-\tau) \, \hat{{\rho}}'_S (t) \,  \sigma^{-}_A \nonumber \\
&+  \alpha_{AD}(0, -\tau) \sigma^{-}_A(-\tau) \, \hat{{\rho}}'_S (t) \,  \sigma^{+}_D  + \alpha_{AA}(0, -\tau) \tilde{\sigma}^{-}_A(-\tau) \, \hat{{\rho}}'_S (t) \, \sigma^{+}_A \Big]  \nonumber \\
&+ \hat{{\rho}}'_S (t)\, \Big [ \alpha_{DD}(-\tau, 0) \tilde{\sigma}^{+}_D(-\tau) \sigma^{-}_D  +  \alpha_{DA} (-\tau, 0)\tilde{\sigma}^{+}_D(-\tau) \sigma^{-}_A \nonumber \\
&+ \bar{\alpha}_{DD}(-\tau, 0)\tilde{\sigma}^{-}_D(-\tau) \sigma^{+}_D  + \bar{\alpha}_{DA}(-\tau, 0) \tilde{\sigma}^{-}_D(-\tau) \sigma^{+}_A \nonumber \\
& +  \alpha_{AD}(-\tau, 0)\tilde{\sigma}^{+}_A(-\tau) \sigma^{-}_D  +  \alpha_{AA} (-\tau, 0)\tilde{\sigma}^{+}_A(-\tau) \sigma^{-}_A \nonumber \\
&+\bar{ \alpha}_{AD}(-\tau, 0) \tilde{\sigma}^{-}_A(-\tau) \sigma^{+}_D  + \bar{\alpha}_{AA} (-\tau, 0)\tilde{\sigma}^{-}_A(-\tau) \sigma^{+}_A \Big]  \Bigg \}
\end{align}
In the above expression, the photon correlation functions $\alpha_{ij}(t, t')$ and $\bar{\alpha}_{ij}(t, t')$ are given by [note that  $\alpha_{ij}(t', t)= \alpha^*_{ij}(t, t')$ and $\bar{\alpha}_{ij}(t', t)=\bar{\alpha}^*_{ij}(t, t')$]:
\begin{align}
\alpha_{ij}(t, t') & \equiv   \langle \hat{\tilde{\mathbb{E}}}(\vec{r}_i, t) \hat{\tilde{\mathbb{E}}}^\dag(\vec{r}_j, t') \rangle  \\
&= \mu^{elec}_i\, \mu^{elec}_j \, e^{i \omega_i t}\, e^{-i \omega_j t'} 
\int_0^\infty d\omega\, \vec{n}^{elec}_i \cdot \int_0^\infty d\omega'\, \langle \vec{E}(\vec{r}_i, \omega) \, \vec{E}^\dag(\vec{r}_j , \omega') \rangle \cdot \vec{n}^{elec}_j \,e^{-i \omega t}\, e^{i \omega' t'}\nonumber \\
& = \frac{\hbar \mu^{elec}_i \mu^{elec}_j}{\epsilon_0 c^2}  e^{i\omega_i t} e^{-i\omega_jt'} \int_0^\infty d\omega\,  \vec{n}^{elec}_i \cdot \mbox{Im} [\tensor{G}(\vec{r}_i, \vec{r}_j, \omega)] \cdot \vec{n}^{elec}_j\, \omega^2 [1+\bar{n}(\omega)]\,e^{-i \omega(t-t')}  \\
\bar{\alpha}_{ij} (t, t') &\equiv  \langle \hat{\tilde{\mathbb{E}}}^\dag(\vec{r}_i, t) \hat{\tilde{\mathbb{E}}}(\vec{r}_j, t') \rangle \\
&= \mu^{elec}_i\, \mu^{elec}_j \, e^{-i \omega_i t}\, e^{i \omega_j t'} 
\int_0^\infty d\omega\, \vec{n}^{elec}_i  \cdot \int_0^\infty d\omega'\, \langle \vec{E}^\dag(\vec{r}_i, \omega) \, \vec{E}(\vec{r}_j , \omega') \rangle \cdot \vec{n}^{elec}_j \,e^{i \omega t}\, e^{-i \omega' t'}\nonumber \\
& = \frac{\hbar \mu^{elec}_i \mu^{elec}_j}{\epsilon_0 c^2}  e^{-i\omega_i t} e^{i \omega_j t'} \int_0^\infty d\omega\, \vec{n}^{elec}_i \cdot \mbox{Im} [\tensor{G}(\vec{r}_i, \vec{r}_j, \omega)] \cdot \vec{n}^{elec}_j\, \omega^2 \bar{n}(\omega)\,e^{i \omega(t-t')}\,.
\end{align}
Here we have used the following relation:
\begin{align}
\vec{\tilde{f}}^{(\dag)}(\vec{r}, \omega) \equiv e^{i \hat{H}_{ph} t}\, \vec{f}^{(\dag)}(\vec{r}, \omega) \, e^{-i \hat{H}_{ph}t} =  e^{\mp i \omega t} \vec{f}^{(\dag)}(\vec{r}, \omega)\,.
\end{align}
Finally, we expand Eq.~\ref{h33} and obtain the following expression:
\begin{align}
(III) &= \Big [\beta^{+-}_{DD}(0, -\tau) \alpha_{DD}(0, -\tau) {\sigma}^{+}_D \tilde{\sigma}^{-}_D(-\tau)  + \beta^{-+}_{DD}(0, -\tau) \bar{\alpha}_{DD} (0, -\tau) {\sigma}^{-}_D \tilde{\sigma}^{+}_D(-\tau)  \nonumber \\
& + \beta^{+-}_{AA}(0, -\tau) \alpha_{AA} (0, -\tau) {\sigma}^{+}_A \tilde{\sigma}^{-}_A(-\tau) +  \beta^{-+}_{AA}(0, -\tau) \bar{\alpha}_{AA} (0, -\tau){\sigma}^{-}_A\tilde{\sigma}^{+}_A(-\tau)\Big] \, \hat{{\rho}}'_S (t) \nonumber \\
&- \Big [\beta^{-+}_{DD}(-\tau, 0) \bar{\alpha}_{DD}(-\tau, 0){\sigma}^{+}_D \, \hat{{\rho}}'_S (t) \, \tilde{\sigma}^{-}_D(-\tau)  + \beta^{+-}_{DD}(-\tau, 0) \alpha_{DD}(-\tau, 0){\sigma}^{-}_D \, \hat{{\rho}}'_S (t) \,  \tilde{\sigma}^{+}_D(-\tau)  \nonumber \\
&+ \beta^{-+}_{AA}(-\tau, 0) \bar{\alpha}_{AA}(-\tau, 0) {\sigma}^{+}_A \, \hat{{\rho}}'_S (t) \,  \tilde{\sigma}^{-}_A (-\tau)+ \beta^{+-}_{AA}(-\tau, 0) \alpha_{AA}(-\tau, 0) {\sigma}^{-}_A \, \hat{{\rho}}'_S (t) \, \tilde{\sigma}^{+}_A(-\tau) \Big] \nonumber \\
&-\Big [\beta^{-+}_{DD}(0, -\tau) \bar{\alpha}_{DD} (0, -\tau) \tilde{\sigma}^{+}_D(-\tau) \, \hat{{\rho}}'_S (t) \,  {\sigma}^{-}_D  + \beta^{+-}_{DD}(0, -\tau) \alpha_{DD}(0, -\tau)\tilde{\sigma}^{-}_D(-\tau)  \, \hat{{\rho}}'_S (t) \, {\sigma}^{+}_D \nonumber \\
& + \beta^{-+}_{AA}(0, -\tau) \bar{\alpha}_{AA} (0, -\tau) \tilde{\sigma}^{+}_A(-\tau) \, \hat{{\rho}}'_S (t) \,  {\sigma}^{-}_A + \beta^{+-}_{AA}(0, -\tau) \alpha_{AA} (0, -\tau)\tilde{\sigma}^{-}_A(-\tau) \, \hat{{\rho}}'_S (t) \, {\sigma}^{+}_A \Big]  \nonumber \\
&+ \hat{{\rho}}'_S (t)\, \Big [\beta^{+-}_{DD}(-\tau, 0) \alpha_{DD} (-\tau, 0) \tilde{\sigma}^{+}_D(-\tau) {\sigma}^{-}_D  + \beta^{-+}_{DD}(-\tau, 0) \bar{\alpha}_{DD}(-\tau, 0)\tilde{\sigma}^{-}_D(-\tau) {\sigma}^{+}_D \nonumber \\
& + \beta^{+-}_{AA}(-\tau, 0) \alpha_{AA}(-\tau, 0) \tilde{\sigma}^{+}_A(-\tau) {\sigma}^{-}_A  + \beta^{-+}_{AA}(-\tau,0 ) \bar{\alpha}_{AA}(-\tau, 0) \tilde{\sigma}^{-}_A(-\tau) {\sigma}^{+}_A \Big]\,,
\end{align}
where the phonons correlation functions $\beta^{\pm \mp}_{ij}(t, t')$ [note that $\beta^{\pm \mp}_{ij}(t', t) = \beta^{\pm \mp *}_{ij}(t, t')$] are given by: 
\begin{align}
\beta^{\pm \mp}_{ij}(t, t') \equiv \langle \hat{\tilde{B}}^{\pm}_{0i}(t) \hat{\tilde{B}}^{\mp}_{0j}(t') \rangle = \langle \hat{B} \rangle^2 (e^{\phi(t-t')} -1)
\end{align}

\bibliographystyle{aipnum4-1}

%

\end{document}